\documentclass[twocolumn,english,superscriptaddress,citeautoscript,showpacs,preprintnumbers,amsmath,amssymb,prd,floatfix,nofootinbib]{revtex4-1}
\pdfoutput=1
\usepackage{amsmath, amsthm, amssymb, graphicx,bm,amstext, mathdots}
\usepackage{hyperref}

\begin{document}
\title{Comment on ``Hiding the Cosmological Constant"}
\author{Qingdi Wang}
\author{William G. Unruh}
\affiliation{Department of Physics and Astronomy, 
The University of British Columbia,
Vancouver, Canada V6T 1Z1}
\maketitle

In a recent Letter \cite{PhysRevLett.123.131302} Carlip argues that the fluctuations in the metric at Planck scales (Wheeler's spactime foam) make it possible to hide the effect of a large cosmological constant. We certainly agree with him that the fluctuations in the metric must be taken into account, and have previously suggested how this might come about \cite{PhysRevD.95.103504, PhysRevLett.125.051301, PhysRevD.102.023537}. However, unfortunately his proposal suffers from some problems. 

His proposal is that the cosmological constant can be hidden because one can find a spacetime foliation by slices of vanishing average expansion. Technically, he starts from an initial spatial slice $\Sigma$ with $\langle K\rangle=0$ and chooses $N, \dot{N}, \ddot{N}$, etc on $\Sigma$ to make all the time derivatives $\frac{d^n\langle K\rangle}{dt^n}\Big|_\Sigma=0$. In this way, he finds a family of spatial slices $\{t=\mathrm{constants}\}$ for which $\langle K\rangle\equiv 0$ (at least for short times).

However, $\langle K\rangle\equiv 0$ does not mean that the macroscopic volume $V_{\mathcal{U}}=\int_{\mathcal{U}}\sqrt{g}d^3x$ he looks at stays constant. Direct calculation gives
\begin{eqnarray}
\frac{dV_{\mathcal{U}}}{dt}\bigg|_\Sigma&=&\int_{\mathcal{U}}NK\sqrt{g}d^3x\neq \left(\langle K\rangle V_{\mathcal{U}}\right)\bigg|_\Sigma=0,\label{1}\\
\frac{d^2V_{\mathcal{U}}}{dt^2}\bigg|_\Sigma&=&\int_\mathcal{U}\left(\dot{N}K+N^2\left(-R+3\Lambda\right)+ND^iD_iN\right)\sqrt{g}d^3x\nonumber\\
&\neq&\left(\frac{d\langle K\rangle}{dt}V_{\mathcal{U}}\right)\bigg|_\Sigma=0,\label{2}
\end{eqnarray}
which mean that $V_{\mathcal{U}}$ is certainly not constant with $t$.

Moreover, a choice of $N$ corresponds to a choice of coordinates, and no physics can depend purely on the choice of coordinates. If distances between geodesics, or more importantly, the wavelengths of fields, grow with time (usually proper time, not coordinate time) they will do so in all coordinate systems, and cannot be hidden by a coordinate choice. One should not try to hide the cosmological constant by choosing $N$ in the first place.

Let us first take a look at how he chooses $N$. We can see from Eq.(7) in \cite{PhysRevLett.123.131302} that for the points in $A=\{x\in\Sigma:R(x)<3\Lambda\}$ we have $\mathcal{L}_n(K\sqrt{g})>0$, which means that they are accelerating toward expansion while for the points in $B=\{x\in\Sigma:R(x)>3\Lambda\}$ we have $\mathcal{L}_n(K\sqrt{g})<0$, which means they are accelerating toward contraction\footnote{There is an extra term $D^iD_iN$ which was omitted by Carlip (see Eq.(5) in \cite{Wang:2019longcomment}). Carlip assumed that it is zero (or at least its average is zero). If put this term back, then $A=\{x\in\Sigma:R(x)<3\Lambda+D^iD_iN\}$, $B=\{x\in\Sigma:R(x)>3\Lambda+D^iD_iN\}$ and our argument does not change.}. Since the average spatial curvature $\langle R\rangle=0$, the volume of $A$ is larger (smaller) than the volume of $B$ if $\Lambda>0$ $(\Lambda<0)$. In order to make $d\langle K\rangle/dt=0$, one has to choose $N$ to be small (large) at points in $A$ and large (small) at points in $B$. Then for the same amount of coordinate time change $\delta t$, the proper time changes $\delta\tau=N\delta t$ are small (large) at points in $A$ and large (small) at points in $B$. The physical progress of time is measured by proper time instead of coordinate time, so what Carlip is really doing is trying to keep $\langle K\rangle=0$ by slowing down (speeding up) the progress of time in places that are accelerating toward expansion and speeding up (slowing down) the progress of time in places that are accelerating toward contraction. To say that this hides the effect of $\Lambda$ is like saying that if you delay the test in places where there are infections, then the spread of the coronavirus is slowed down. Of course it does not work. What is really happening is that the overall space expands (contracts) at an accelerated rate.

This can also be seen from a different perspective. Note that $V_{\mathcal{U}}$ represents the volume occupied by the particles enclosed in the region $\mathcal{U}$ whose world lines are orthogonal to the slices $\{t=\mathrm{constants}\}$. If choose $N=1$, the world lines orthogonal to these slices are geodesics with proper time $\tau=t$ and $V_{\mathcal{U}}$ represents the volume occupied by these geodesic particles. Using the second line of Eq.(7) in \cite{PhysRevLett.123.131302} we obtain
\begin{equation}\label{geodesic average}
\frac{d\langle K\rangle}{d\tau}\bigg|_\Sigma=3\Lambda,
\end{equation}
where we have used the requirement that $\langle R\rangle=0$. Then \eqref{1} and \eqref{2} become
\begin{eqnarray}
\frac{dV_{\mathcal{U}}}{d\tau}\bigg|_\Sigma&=&\left(\langle K\rangle V_{\mathcal{U}}\right)\bigg|_\Sigma =0 \label{4}\\
\frac{d^2V_{\mathcal{U}}}{d\tau^2}\bigg|_\Sigma&=&\left(\frac{d\langle K\rangle}{d\tau}V_{\mathcal{U}}\right)\bigg|_\Sigma=3\Lambda V_{\mathcal{U}}\bigg|_\Sigma. \label{5}
\end{eqnarray}
Therefore, the geodesic volume $V_{\mathcal{U}}$ will disastrously explode if $\Lambda>0$ or collapse if $\Lambda<0$.

The behavior of geodesic particles who feels only gravity reveals the gravitational property of $\Lambda$. If $N$ is position dependent, the particles following world lines orthogonal to the slices $\{t=\mathrm{constants}\}$ are non-geodesics and there are nongravitational forces $\mathbf{F}=m\mathbf{D}N/N$ acting on them (see Eq.(3.17) in \cite{Gourgoulhon:2007ue}). Carlip thinks that the nongeodesic volume who satisfies \eqref{1}, \eqref{2} does not explode or collapse violently like the geodesic volume who satisfies \eqref{4}, \eqref{5} because the effect of the large cosmological constant $\Lambda$ is hidden. However, it is simply because those nongravitational forces are trying to balance the effect of $\Lambda$.

In summary, the effect of $\Lambda$ is not hidden in the way proposed in \cite{PhysRevLett.123.131302}. Fortunately, this is not the end of the story. It can be hidden in Planck scale curvature fluctuations in a different way \cite{PhysRevLett.125.051301, PhysRevD.102.023537, Wang:2019longcomment}.

Again, we emphasize that we agree completely with Carlip that one must take the fluctuations of curvature at small scales seriously, and that simply assuming a homogeneous and isotropic spacetime is not sufficient.

\bibliographystyle{unsrt}
\bibliography{how_vacuum_gravitates}

\end{document}